\documentclass[twocolumn,pra,longbibliography,footinbib]{revtex4-1}
\usepackage{amssymb}
\usepackage{amsmath}
\usepackage{dcolumn}
\usepackage{bm}
\usepackage{graphicx}
\usepackage{subfigure}
\usepackage{mathrsfs}
\usepackage{color}
\usepackage{url}
\usepackage{lipsum}
\usepackage[colorlinks]{hyperref}

\hypersetup{ plainpages=true,
 breaklinks=true,        hypertexnames=false,   pageanchor=true,
 colorlinks=true,
 linkcolor={blue},
 citecolor={blue},
 urlcolor={blue},
  anchorcolor={black}
}

\begin{document}

\title{Chiral cavity-magnonic system for the unidirectional emission of a tunable squeezed microwave field}
\author{Ji-kun Xie}
\author{Sheng-li Ma}
\email{msl1987@xjtu.edu.cn}
\author{Ya-long Ren}
\author{Shao-yan Gao}
\author{Fu-li Li}

\address{
MOE Key Laboratory for Non-equilibrium Synthesis
and Modulation of Condensed Matter, Shaanxi Province Key Laboratory
of Quantum Information and Quantum Optoelectronic Devices, School of
Physics, Xi’an Jiaotong University, Xi’an 710049, China}

\begin{abstract}

Unidirectional photon emission is crucial for constructing quantum networks and realizing scalable quantum
information processing. In the present work an efficient scheme is developed for the unidirectional emission of
a tunable squeezed microwave field. Our scheme is based on a chiral cavity magnonic system, where a magnon
mode in a single-crystalline yttrium iron garnet (YIG) sphere is selectively coupled to one of the two degenerate
rotating microwave modes in a torus-shaped cavity with the same chirality. With the YIG sphere driven by a
two-color Floquet field to induce sidebands in the magnon-photon coupling, we show that the unidirectional
emission of a tunable squeezed microwave field can be generated via the assistance of the dissipative magnon
mode and a waveguide. Moreover, the direction of the proposed one-way emitter can be controlled on demand by reversing the biased magnetic field. Our work opens up an avenue to create and manipulate one-way nonclassical
microwave radiation field and could find potential quantum technological applications.

\end{abstract}

\maketitle

\section{Introduction}

Chiral quantum optics has become a burgeoning field due to its potential
applications in quantum information processing and quantum networks \cite%
{Coptic1,Coptic2,Coptic3,Coptic4,Coptic6}. It seeks to exploit new
approaches and systems exhibiting chiral light-matter interactions \cite%
{CP1,CP2,CP3,CP4,CP5,CP6,CP7,CP8,CP9,CP10,CP11}. On this subject,
nanophotonic systems, such as nanoscale waveguides and
whispering-gallery-mode (WGM) resonators \cite{nano1,nano2,nano3}, have
emerged as attractive candidates, where the light field is strongly
transversely confined in a subwavelength space and exhibits the optical
spin-orbit coupling \cite{SO,SO1}. The chiral interaction can be achieved by
coupling the spin-momentum-locked light to quantum emitters with
polarization-dependent dipole transitions \cite%
{Ccouple1,Ccouple2,Ccouple3,Ccouple4,Ccouple5}. And a large number of chiral
devices have been proposed theoretically and demonstrated experimentally,
like single-photon diodes \cite{diode1,circu1}, single-photon routing \cite%
{route1,route2,route3}, circulators \cite{circu1,circu2}, isolators \cite%
{circu2,iso0,iso1} and etc. These elements are key components for building
large-scale quantum networks.

In parallel, the chiral light-matter interactions can also be utilized to
control the direction of photon emission \cite%
{Cemission1,Cemission2,Cemission3,Cemission4}. By designing the suitable
chiral coupling between quantum emitters and evanescent fields, the
deterministic and the highly directional photon emission can be achieved
along only one of the selected directions \cite{nano1,Ccouple1,Cemission4}.
Recently, the chiral-waveguide-based and the chiral-cavity-based systems
have been explored to generate the unidirectional emission of single photons
\cite{single1,single2,single3,single4,single5}. In addition, the
unidirectional laser emission has also been demonstrated through the
non-Hermitian scattering induced chirality in the WGM resonator \cite%
{laser1,laser2}. The unidirectional photon emission can facilitate the
free-space coupling of energy, improve the collection efficiency of weak
optical signals and on-demand distribute quantum information \cite%
{eff1,eff2,eff3,eff4,eff5}.

In recent years, the hybrid quantum system of yttrium iron garnet (YIG) and
superconducting circuit has been considered as a powerful platform for
quantum information applications \cite{hybrid1,hybrid2,hybrid3,hybrid4}. Due
to the high spin density and low damping rate of the ferrimagnetic insulator
YIG, the strong and even ultrastrong coupling between a magnetostatic mode
in a YIG sphere and a microwave cavity mode have been experimentally
observed \cite{strong1,strong2,strong3,strong4,strong5,strong6}. Many
intriguing phenomena have been studied in this hybrid system, such as the
generation of various quantum states \cite%
{state0,state1,state2,state3,state4,state5,state6,state7,state8,state9,state10,state11},
nonreciprocity \cite{nonre1,nonre2,nonre3}, non-Hermitian physics \cite%
{nonH1,nonH2,nonH3}, and Floquet engineering \cite{Floq1,Floq2,Floq3,Floq4}.
Remarkably, magnons also carry angular momentum or \textquotedblleft
spin\textquotedblright. For instance, the Kittel mode magnetization
precesses counterclockwise around the applied magnetic field \cite{SOG}.
Analogous with the chiral coupling of spin-polarized atoms or quantum dots
with spin-momentum-locked light \cite{nano1,nano2,nano3}, the Kittel mode
can only couple to photons with the same polarization in ring-shaped cavity
or waveguide \cite{couple0,couple1,couple2,couple3,couple4}. Therefore, this
hybrid magnonic system can act as a promising architecture for studying
chiral quantum optics \cite{chiral1,chiral2,chiral3,chiral4,chiral5}.

\begin{figure}[tbh]
\centering \includegraphics[width=8cm]{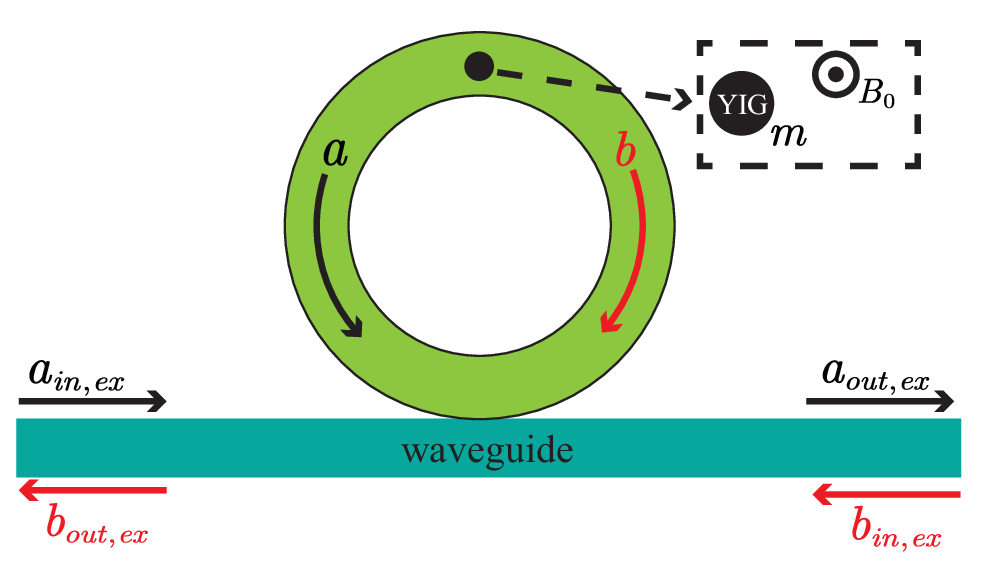}
\caption{(Color online) Schematic diagram of the hybrid quantum setup. A
torus-shaped cavity supports both CCW ($a$) and CW ($b$) rotating microwave
modes, which are evanescently coupled to a nearby microwave waveguide. A
small ferromagnetic YIG sphere is placed inside the cavity and biased
perpendicularly by a static magnetic field $B_{0}$ to establish the
magnon-photon coupling. Additionally, the magnon mode is driven by a
two-color Floquet driving field (not shown) along the bias field direction.}
\end{figure}
In this paper, we develop a method for the unidirectional emission of a
tunable squeezed microwave field based on a chiral cavity magnonic system.
In our scheme, a torus-shaped microwave cavity that supports a pair of
degenerate counter-rotating microwave modes is exploited to chirally coupled
to a YIG sphere. The chiral interaction can be achieved by locating the YIG
sphere on a special chiral line in the cavity, where the circular
polarization of the cavity mode is locked to its propagation direction \cite%
{couple2,couple3,couple4,chiral1,chiral2}. As a result, the Kittel mode will
only couple to one of the microwave modes with the same polarization. By
applying a two-color Floquet driving field to the magnetic sphere to induce
sideband transitions, we can engineer a squeezing-type interaction between
the magnon mode and the selected microwave mode, which can be steered into a
squeezed state with the aid of large decay rate of the magnon mode. By
further considering the cavity evanescently coupled to a microwave
waveguide, we can generate a unidirectionally squeezed microwave source.
Notably, the emission direction can be conveniently adjusted by reversing
the bias magnetic field, and the amount of squeezing is tunable by changing
the external parameters. Moreover, the numerical simulations indicate that
the unidirectional emission is robust against the imperfect chiral coupling.
Our work could find a wide range of practical applications in quantum networks
and quantum sensing.

\section{Model}

As schematically shown in Fig. 1, the hybrid cavity magnonic system under
consideration consists of a torus-shaped microwave cavity with a small
highly polished YIG sphere placed inside \cite{couple3,couple4}. Due to the
geometrical rotational symmetry, the cavity supports a pair of degenerate
counterclockwise (CCW) and clockwise (CW) microwave modes, which are
evanescently coupled to a nearby microwave waveguide. Under a static
out-of-plane bias magnetic field at a strength $B_{0}$, many magnetostatic
modes will be excited in the YIG sphere \cite{strong1,strong2,strong3}. In
our scheme, we focus only on the Kittel mode, which has the uniform spin
precessions in the whole volume of the magnetic sphere. The two microwave
modes are coupled to the Kittel mode through the collective magnetic-dipole
interaction, respectively. Now, we can give the free Hamiltonian of the
Kittel mode and the cavity modes (we set $\hbar =1$ hereafter)%
\begin{equation}
H_{0}=\omega _{m}m^{\dag }m+\omega _{0}\sum_{\alpha =a,b}\alpha ^{\dag
}\alpha .  \label{eq1}
\end{equation}%
Here, $m$ $(m^{\dag })$ is the boson operator of the magnon mode with the
resonance frequency $\omega _{m}=\gamma B_{0}$, where $\gamma =2\pi \times
28 $ GHz/T is the gyromagnetic ratio. $a$ $(a^{\dag })$ and $b$ $(b^{\dag })$
represent the annihilation (creation) operators for the CCW and CW microwave
modes with the resonance frequency $\omega _{0}$.

We then consider the chiral photon-magnon interaction, which takes the form \cite{couple2,couple3,couple4,chiral1,chiral2}
\begin{equation}
H_{int}=\sum_{\alpha =a,b}g_{\alpha }(\alpha ^{\dag }+\alpha )(m^{\dag }+m)
\label{eq2}
\end{equation}%
with $g_{\alpha }(\alpha =a,b)$ describing the coupling strength between the
cavity mode $\alpha $ and the magnon mode $m$. To be specific, the Kittel
mode magnetization precesses around the effective magnetic field in an
anticlockwise manner and couples preferentially to photons with the same
polarization. Here, the magnetic field of each TE mode is
transversally confined in the cavity, so that it has a strong longitudinal
polarization component along the propagation direction \cite%
{Coptic1,route1,longitu}. Moreover, the longitudinal and transverse
polarization components oscillate $\pm 90^{\circ }$ out of phase with each
other, where $+$ ($-$) sign depends on the propagation direction of the
cavity modes. At the special radial locations, the magnetic field can be
circularly polarized when the longitudinal and transverse polarization
components have the same amplitudes. Also, since the polarization direction
is locked to the sign of their linear momentum, the counterpropagated CW and
CCW modes can thereby possess mutually orthogonal polarizations and opposite
chiralities \cite{couple2}. As a result, the magnon mode can only couple to
one of them with the same chirality, i.e., $g_{a}\neq 0$ and $g_{b}=0$, or $%
g_{b}\neq 0$ and $g_{a}=0$.

In addition, the chiral coupling can be well
controlled by reversing the direction of the biased magnetic field, because
it is accompanied with the reversion of the precession of the Kittel mode.
Alternatively, each of the two degenerate counter-propagating microwave
modes can have opposite chirality at different radial locations \cite%
{couple3,couple4}, so the chiral coupling can also be controlled by shifting
the magnet position.

Furthermore, a two-color Floquet driving field is applied to the YIG sphere
along the bias magnetic field direction, under which the oscillating
frequency of the magnon mode is modulated accordingly. So the coupling
between the Floquet driving field and the Kittel mode is described by \cite{Floq1,Floq2,Floq3,Floq4}
\begin{equation}
H_{d}(t)=[\Omega _{1}\cos (\omega _{d1}t)+\Omega _{2}\cos (\omega
_{d2}t+\theta )]m^{\dag }m,  \label{eq3}
\end{equation}%
where $\Omega _{j}$ $(j=1,2)$ denotes the driving amplitudes with the driving
frequencies $\omega _{dj}$, $\theta $ is the relative phase. Experimentally, this manipulation can be
implemented through a small coil being looped tightly around the magnetic
sphere to modulate the bias magnetic field \cite{Floq1}. Under this kind of
periodical modulation, the desired parametric magnon-photon interaction can
be sculpted to realize the one-way squeezed microwave source.

\begin{figure}[tbh]
\centering \includegraphics[width=8cm]{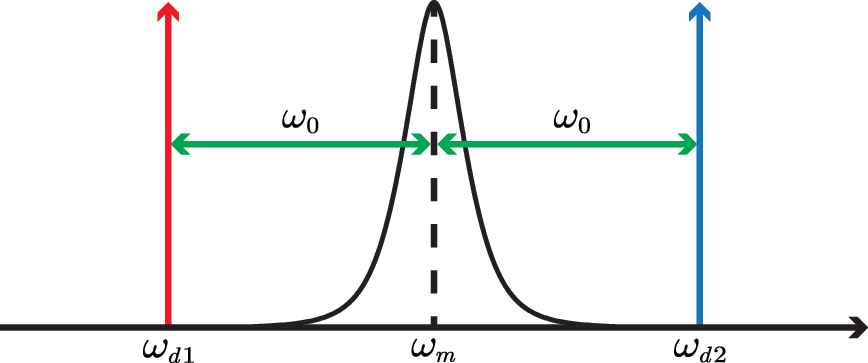}
\caption{(Color online) Schematic diagram of sideband transitions of the coupled magnon-photon modes, where $\protect\omega _{d1}=\protect\omega _{m}-\protect\omega _{0}$
and $\protect\omega _{d2}=\protect\omega _{m}+\protect\omega _{0}$
correspond to the red and blue sidebands, respectively.}
\end{figure}

\section{Unidirectional emission of a tunable squeezed microwave field}

In this section, we will detail the procedure for unidirectionally creating
a tunable squeezed microwave field based on above introduced chiral cavity
magnonic system. The chirality stems from the selective coupling of the
magnon mode to one of the two degenerate rotating microwave modes with the
same polarization. Combining with the Floquet driving to engineer a desired
squeezing-type interaction, we can generate a chiral squeezed microwave
radiation under the assistance of a dissipative magnon mode and a waveguide.

\subsection{The effective Hamiltonian for the generation of a chiral
squeezed state}

We start to derive the effective Hamiltonian to selectively generate a chiral
squeezed state; that is, only one of the two cavity modes will be squeezed. To this end, the Floquet
driving to the magnon mode plays a key role. We define the following rotating transformation:%
\begin{eqnarray}
U(t) &=&\mathcal{T}\exp \{-i\int_{0}^{t}[H_{0}+H_{d}(\tau )]d\tau \}  \notag
\\
&=&\exp \{-i(\omega _{m}m^{\dag }m+\omega _{0}\sum_{\alpha =a,b}\alpha
^{\dag }\alpha )t  \notag \\
&&-i[\xi _{1}\sin (\omega _{d1}t)+\xi _{2}\sin (\omega _{d2}t+\theta
)]m^{\dag }m\},  \label{eq4}
\end{eqnarray}%
where $\mathcal{T}$ is the time order operator and $\xi _{j}=\Omega
_{j}/\omega _{dj}(j=1,2)$. In the rotating frame with respect to $U(t)$, the total
Hamiltonian $H_{t}=H_{0}+H_{int}+H_{d}(t)$ of the system will become%
\begin{eqnarray}
H_{I} &=&U^{\dag }(t)H_{t}U(t)+i\frac{dU^{\dag }(t)}{dt}U(t)  \notag \\
&=&\sum_{\alpha =a,b}g_{\alpha }m^{\dag }e^{i[\omega _{m}t+\xi _{1}\sin
(\omega _{d1}t)+\xi _{2}\sin (\omega _{d2}t+\theta )]}  \notag \\
&&\times (\alpha e^{-i\omega _{0}t}+\alpha ^{\dag }e^{i\omega _{0}t})+H.c.  \label{eq5}
\end{eqnarray}

By using the Jacobi-Anger identity%
\begin{equation}
e^{ix\sin \phi }=\sum_{n=-\infty }^{n=\infty }J_{n}(x)e^{in\phi }  \label{eq6}
\end{equation}%
with the $n$th Bessel functions of the first kind $J_{n}(x)$, the
Hamiltonian can be rewritten as%
\begin{eqnarray}
H_{I} &=&\sum_{\alpha =a,b}g_{\alpha }m^{\dag }e^{i\omega
_{m}t}\sum_{n=-\infty }^{\infty }J_{n}(\xi _{1})e^{in\omega
_{d1}t}\sum_{k=-\infty }^{\infty }J_{k}(\xi _{2})  \notag \\
&&\times e^{ik(\omega _{d2}t+\theta )}(\alpha e^{-i\omega _{0}t}+\alpha
^{\dag }e^{i\omega _{0}t})+H.c.  \label{eq7}
\end{eqnarray}%

To produce the desired magnon-photon coupling, as displayed in Fig. 2, we choose the Floquet driving
frequencies satisfying $\omega _{d1}=\omega _{m}-\omega _{0}$ and $\omega
_{d2}=\omega _{m}+\omega _{0}$, which correspond to the red and blue sideband
transitions, respectively. Furthermore, under the condition $%
g_{\alpha }\ll \{\omega _{dj},\omega _{m},\omega _{0}\}$, we can only keep
the resonant terms in Eq. (\ref{eq7}), but safely discard those fast
oscillating terms by means of the rotating-wave approximation. Then, we can
obtain the effective Hamiltonian of the system
\begin{equation}
H_{eff}=\sum_{\alpha =a,b}G_{\alpha }m^{\dag }(\alpha +\varepsilon
e^{-i\theta }\alpha ^{\dag })+H.c.,  \label{eq8}
\end{equation}%
where we have $G_{\alpha }=g_{\alpha }J_{-1}(\xi _{1})J_{0}(\xi _{2})$, and $%
\varepsilon e^{-i\theta } =J_{0}(\xi _{1})J_{-1}(\xi _{2})e^{-i\theta }/J_{-1}(\xi _{1})J_{0}(\xi _{2})$
is the ratio of the parametric-amplifier interaction to the
Jaynes-Cummings-type coupling. In what follows, we exclusively
consider the situation of $\left\vert \varepsilon \right\vert <1$, which
ensures that Eq. (\ref{eq8}) is stable. It is now clear that $H_{eff}$
describes a squeezing-type interaction, and can be used to produce the
squeezed state. However, due to the chiral coupling, i.e., $G_{a}\neq 0$ and
$G_{b}=0$, or $G_{b}\neq 0$ and $G_{a}=0$, we can only generate a squeezed state in one of the two cavity modes.

By introducing a hybridized mode $A=(G_{a}a+G_{b}b)/\sqrt{%
G_{a}^{2}+G_{b}^{2}}$, the Hamiltonian in Eq. (\ref{eq8}) can be rewritten as%
\begin{equation}
H_{eff}=\sqrt{G_{a}^{2}+G_{b}^{2}}[m^{\dag }(A+\varepsilon e^{-i\theta
}A^{\dag })+H.c.].  \label{eq9}
\end{equation}%
To clearly reveal the mechanism for generating the chiral squeezed state, we
perform the unitary squeezing transformation $H_{S}=S_{A}^{\dag }(\zeta
)H_{eff}S_{A}(\zeta )$ \cite{Bog}, where $S_{A}(\zeta )=\exp [(\zeta
A^{2}-\zeta ^{\ast }A^{\dag 2})/2]$ represents the single-mode squeezing
operator with $\zeta =re^{i\theta }$. Here $r$ is squeezing parameter
defined by $\tanh r=\varepsilon $ and $\theta $ denotes the squeezing angle,
both of which can be controlled on-demand by adjusting the two-color Floquet
driving field. {In the squeezed frame, the effective Hamiltonian (\ref{eq9}) is
transformed to the form
\begin{equation}
H_{S}=\sqrt{\left( G_{a}^{2}+G_{b}^{2}\right) \left( 1-\varepsilon
^{2}\right) }(m^{\dag }A+mA^{\dag }).  \label{eq10}
\end{equation}%
Obviously, $H_{S}$ describes a beam-splitter interaction. If we
set $G_{a}\neq 0$ and $G_{b}=0$ ($G_{b}\neq 0$ and $G_{a}=0$) via the
selective coupling rule, the mode $A=a$ ($A=b$) can be cooled down to the
vacuum state in the squeezed picture via the dissipation of the magnon mode.
Reversing the squeezing transformation, the cavity mode $a$ ($b$) is
actually steered into a single-mode squeezed state. In this way, a chiral
squeezed state is created. By further considering the microwave cavity
coupled to a waveguide, the squeezed output field can be emitted along the
right (left) direction of the waveguide. This is the basic idea for the
unidirectional emission of a squeezed source.

\subsection{Unidirectional squeezing emission}

We now study the unidirectional squeezing emission by considering the
coupling of the microwave cavity coupled to a transmission line that acts
the role of a waveguide. According to the Hamiltonian (\ref{eq8}) and the
standard input--output theory, the general quantum Langevin equations of the
system is given by
\begin{subequations}
\begin{eqnarray}
\dot{\alpha} &=&-\frac{\kappa }{2}\alpha -iG_{\alpha }(m+\varepsilon
e^{-i\theta }m^{\dag })+\sqrt{\kappa _{ex}}\alpha _{in,ex}  \notag \\
&&+\sqrt{\kappa _{0}}\alpha _{in,0},  \label{eq11a}
\end{eqnarray}%
\begin{equation}
\dot{m}=-\frac{\gamma _{m}}{2}m-i\sum_{\alpha =a,b}G_{\alpha }(\alpha
+\varepsilon e^{-i\theta }\alpha ^{\dag })+\sqrt{\gamma _{m}}m_{in}.  \label{eq11b}
\end{equation}%
Without loss of generality, the cavity modes $a$ and $b$ are assumed to have
the same damping rate $\kappa =\kappa _{0}+\kappa _{ex}$, where $\kappa _{0}$
is the intrinsic decay rate, and $\kappa _{ex}$ denotes the external
coupling between the cavity mode and the waveguide. In addition, $\alpha
_{in,0}$ and $\alpha _{in,ex}$ $(\alpha =a,b)$ represent the noise
operators. $m_{in}$ is the noise operator of the magnon mode, and $\gamma
_{m}$ is the associated damping rate. These noise operators obey the
following correlation relations
\end{subequations}
\begin{subequations}
\begin{equation}
\left\langle \alpha _{in,ex}(t)\alpha _{in,ex}^{\dag }(t^{\prime
})\right\rangle =\delta (t-t^{\prime }),  \label{eq12a}
\end{equation}%
\begin{equation}
\left\langle \alpha _{in,0}(t)\alpha _{in,0}^{\dag }(t^{\prime
})\right\rangle =\delta (t-t^{\prime }),  \label{eq12b}
\end{equation}%
\begin{equation}
\left\langle m_{in}(t)m_{in}^{\dag }(t^{\prime })\right\rangle =\delta
(t-t^{\prime }),  \label{eq12c}
\end{equation}%
where we have neglected the thermal effects. Because, at the experimental
working temperature $T=20$ mK, the average thermal excitation number of a
boson mode with resonance frequency $6.5$ GHz is about $10^{-7}$, which is
thus negligible.

By introducing the Fourier transformation $o(t)=\int_{-\infty }^{+\infty
}o(\omega )e^{-i\omega t}d\omega /2\pi $ for an arbitrary operator $o$, we
can rewrite the Eqs. (\ref{eq11a}) and (\ref{eq11b}) in the frequency domain
as
\end{subequations}
\begin{subequations}
\begin{eqnarray}
\alpha (\omega ) &=&\frac{1}{\frac{\kappa }{2}-i\omega }\{-iG_{\alpha
}[m(\omega )+\varepsilon e^{-i\theta }m^{\dag }(-\omega )]  \notag \\
&&+\sqrt{\kappa _{ex}}\alpha _{in,ex}(\omega )+\sqrt{\kappa _{0}}\alpha
_{in,0}(\omega )\},  \label{eq13a}
\end{eqnarray}%
\begin{eqnarray}
m(\omega ) &=&\frac{1}{\frac{\gamma _{m}}{2}-i\omega }\{-i\sum_{\alpha
=a,b}G_{\alpha }[\alpha (\omega )  \notag \\
&&+\varepsilon e^{-i\theta }\alpha ^{\dag }(-\omega )]+\sqrt{\gamma _{m}}%
m_{in}(\omega )\}.  \label{eq13b}
\end{eqnarray}%
Correspondingly, the correlation functions of Eqs. (\ref{eq12a})-(\ref{eq12c}%
) in the frequency domain yield
\end{subequations}
\begin{subequations}
\begin{equation}
\left\langle \alpha _{in,ex}(\omega )\alpha _{in,ex}^{\dag }(-\omega
^{\prime })\right\rangle =2\pi \delta (\omega +\omega ^{\prime }),
\label{eq14a}
\end{equation}%
\begin{equation}
\left\langle \alpha _{in,0}(\omega )\alpha _{in,0}^{\dag }(-\omega ^{\prime
})\right\rangle =2\pi \delta (\omega +\omega ^{\prime }),  \label{eq14b}
\end{equation}%
\begin{equation}
\left\langle m_{in}(\omega )m_{in}^{\dag }(-\omega ^{\prime })\right\rangle
=2\pi \delta (\omega +\omega ^{\prime }).  \label{eq14c}
\end{equation}%
Our interest is the output fields of the cavity. Hence, in terms of the
standard input-output relation
\end{subequations}
\begin{equation}
\alpha _{out,ex}=\sqrt{\kappa _{ex}}\alpha -\alpha _{in,ex},  \label{eq15}
\end{equation}%
we can derive out the frequency components of the output field operators
\begin{widetext}
\begin{subequations}
\begin{eqnarray}
a_{out,ex}(\omega ) &=&[\frac{N_{b}(\omega )\kappa _{ex}}{D(\omega )}%
-1]a_{in,ex}(\omega )+\frac{1}{D(\omega )}\{N_{b}(\omega )\sqrt{\kappa
_{ex}\kappa _{0}}a_{in,0}(\omega )+8G_{a}G_{b}(1-\varepsilon ^{2})  \notag \\
&&\times \lbrack \kappa _{ex}b_{in,ex}(\omega )+\sqrt{\kappa _{ex}\kappa _{0}%
}b_{in,0}(\omega )]+4iG_{a}\kappa _{-}\sqrt{\kappa _{ex}\gamma _{m}}%
[m_{in}(\omega )+\varepsilon e^{-i\theta }m_{in}^{\dag }(-\omega )]\},
\label{eq16a}
\end{eqnarray}%
\begin{eqnarray}
b_{out,ex}(\omega ) &=&[\frac{N_{a}(\omega )\kappa _{ex}}{D(\omega )}%
-1]b_{in,ex}(\omega )+\frac{1}{D(\omega )}\{N_{a}(\omega )\sqrt{\kappa
_{ex}\kappa _{0}}b_{in,0}(\omega )+8G_{a}G_{b}(1-\varepsilon ^{2})  \notag \\
&&\times \lbrack \kappa _{ex}a_{in,ex}(\omega )+\sqrt{\kappa _{ex}\kappa _{0}%
}a_{in,0}(\omega )]+4iG_{b}\kappa _{-}\sqrt{\kappa _{ex}\gamma _{m}}%
[m_{in}(\omega )+\varepsilon e^{-i\theta }m_{in}^{\dag }(-\omega )]\}
\label{eq16b}
\end{eqnarray}%
\end{subequations}
with $\kappa _{-}=\kappa -2i\omega $, $\gamma _{-}=\gamma _{m}-2i\omega $, $%
N_{\alpha }(\omega )=8G_{\alpha }^{2}(\varepsilon ^{2}-1)-2\kappa _{-}\gamma
_{-}$ and $D(\omega )=4\kappa _{-}(\varepsilon
^{2}-1)(G_{a}^{2}+G_{b}^{2})-\kappa _{-}^{2}\gamma _{-}$.

Now, the spectrum of the
output microwave field for the $\alpha $ mode is given by
\begin{equation}
S_{\alpha ,out}(\omega )=\frac{1}{4\pi }\int_{-\infty }^{+\infty }d\omega
^{\prime }e^{-i(\omega +\omega ^{\prime })t}[\left\langle X_{\alpha
,out}^{\theta }(\omega )X_{\alpha ,out}^{\theta }(\omega ^{\prime
})\right\rangle +\left\langle X_{\alpha ,out}^{\theta }(\omega ^{\prime
})X_{\alpha ,out}^{\theta }(\omega )\right\rangle ],
\label{eq17}
\end{equation}%
where $X_{\alpha ,out}^{\theta }(\omega )=\alpha _{out,ex}(\omega )e^{\frac{%
i\theta }{2}}+\alpha _{out,ex}^{\dag }(-\omega )e^{-\frac{i\theta }{2}}$
is the associated quadrature operator of the output
fields. Substituting Eqs. (\ref{eq16a}) and (\ref{eq16b}) into Eq. (\ref{eq17}%
), we can obtain the squeezing spectra of the output fields
\begin{subequations}
\begin{eqnarray}
S_{a,out}(\omega ) &=&\left\vert \frac{N_{b}(\omega )\kappa _{ex}}{D(\omega )%
}-1\right\vert ^{2}+\left\vert \frac{N_{b}(\omega )\sqrt{\kappa _{ex}\kappa
_{0}}}{D(\omega )}\right\vert ^{2}+\left\vert \frac{8G_{a}G_{b}(1-%
\varepsilon ^{2})\kappa _{ex}}{D(\omega )}\right\vert ^{2}  \notag \\
&&+\left\vert \frac{8G_{a}G_{b}(1-\varepsilon ^{2})\sqrt{\kappa _{ex}\kappa
_{0}}}{D(\omega )}\right\vert ^{2}+\left\vert \frac{4G_{a}\kappa
_{-}(1-\varepsilon )\sqrt{\kappa _{ex}\gamma _{m}}}{D(\omega )}\right\vert
^{2},  \label{eq18a}
\end{eqnarray}%
\begin{eqnarray}
S_{b,out}(\omega ) &=&\left\vert \frac{N_{a}(\omega )\kappa _{ex}}{D(\omega )%
}-1\right\vert ^{2}+\left\vert \frac{N_{a}(\omega )\sqrt{\kappa _{ex}\kappa
_{0}}}{D(\omega )}\right\vert ^{2}+\left\vert \frac{8G_{a}G_{b}(1-%
\varepsilon ^{2})\kappa _{ex}}{D(\omega )}\right\vert ^{2}  \notag \\
&&+\left\vert \frac{8G_{a}G_{b}(1-\varepsilon ^{2})\sqrt{\kappa _{ex}\kappa
_{0}}}{D(\omega )}\right\vert ^{2}+\left\vert \frac{4G_{b}\kappa
_{-}(1-\varepsilon )\sqrt{\kappa _{ex}\gamma _{m}}}{D(\omega )}\right\vert
^{2}.  \label{eq18b}
\end{eqnarray}%
\end{subequations}
\end{widetext}
To quantify the amount of squeezing of the output fields, we now define the
quantum noise reduction $F_{\alpha ,out}(\omega )=-10\log _{10}[S_{\alpha
,out}(\omega )/S_{\alpha ,out}^{vac}(\omega )]$ in dB units with $S_{\alpha
,out}^{vac}(\omega )=1$ being the squeezing spectrum of the vacuum state. Note that if the output field is
squeezed, the quantum noise reduction $F_{\alpha ,out}(\omega )$ is larger
than zero. Clearly, to achieve a perfect one-way squeezing emitter, the
condition $F_{a,out}(\omega )>0$ and $F_{b,out}(\omega )=0$ ($%
F_{b,out}(\omega )>0$ and $F_{a,out}(\omega )=0$) should be satisfied, which
indicates that the squeezed field is only emitted along the right (left)
direction of the waveguide.

To well grasp the main result of this work, we start our
analysis with the ideal chiral coupling, i.e., $g_{a}\neq 0$ and $g_{b}=0$.
Then, the output squeezing spectra in Eqs. (\ref{eq18a}) and (\ref{eq18b}) yield
\begin{subequations}
\begin{eqnarray}
S_{a,out}(\omega ) &=&\left\vert \frac{\kappa _{-}\gamma _{-}(2\kappa
_{ex}-\kappa _{-})-4\kappa _{-}(1-\varepsilon ^{2})G_{a}^{2}}{\kappa
_{-}^{2}\gamma _{-}+4\kappa _{-}(1-\varepsilon ^{2})G_{a}^{2}}\right\vert
^{2}  \notag \\
&&+\left\vert \frac{2\kappa _{-}\gamma _{-}\sqrt{\kappa _{ex}\kappa _{0}}}{%
\kappa _{-}^{2}\gamma _{-}+4\kappa _{-}(1-\varepsilon ^{2})G_{a}^{2}}%
\right\vert ^{2}  \notag \\
&&+\left\vert \frac{4G_{a}\kappa _{-}(\varepsilon -1)\sqrt{\kappa
_{ex}\gamma _{m}}}{\kappa _{-}^{2}\gamma _{-}+4\kappa _{-}(1-\varepsilon
^{2})G_{a}^{2}}\right\vert ^{2},  \label{eq19a} \\
S_{b,out}(\omega ) &=&1,  \label{eq19b}
\end{eqnarray}
\end{subequations}
\begin{figure}[tbh]
\includegraphics[width=9cm]{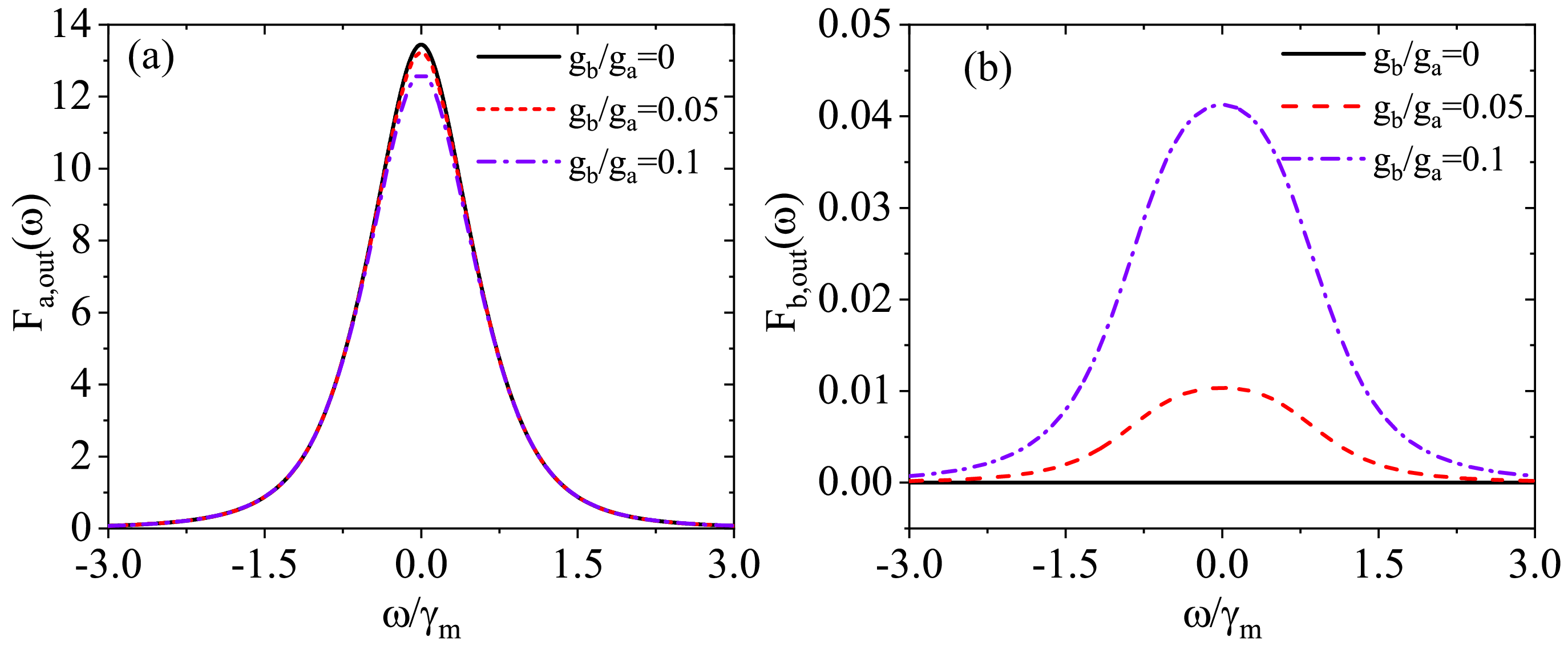}
\caption{(Color online) The squeezing spectra $F_{a,out}(\protect\omega )$
[(a)] and $F_{b,out}(\protect\omega )$ [(b)] versus the dimensionless frequency $%
\protect\omega /\protect\gamma _{m}$ under the different coupling ratios $%
g_{b}/g_{a}=0$, $0.05$, $0.1$. The related parameters are chosen as $G_{a}/%
\protect\gamma _{m}=2.5$, $\protect\varepsilon =0.95$, $\protect\kappa _{0}/%
\protect\gamma _{m}=0.05$, $\protect\kappa _{ex}/\protect\gamma _{m}=2.5$
and $\protect\gamma _{m}/2\protect\pi =2$ MHz.}
\end{figure}
It is now clear that the spectra $S_{a,out}(\omega )<1$ and $%
S_{b,out}(\omega )=1$ can be satisfied under the situation of perfect
chirality, which is equivalent to $F_{a,out}(\omega )>0$ and $%
F_{b,out}(\omega )=0$. So, a desired one-way squeezed source can be created;
that is, the squeezed microwave field can only be emitted in one direction.
To exhibit the unidirectional squeezing emission, we choose $\varepsilon
=0.95$, and show $F_{a,out}(\omega )$ and $F_{b,out}(\omega )$ versus the
dimensionless frequency $\omega /\gamma _{m}$ in Fig. 3. It is observed that
$F_{a,out}(\omega )$ is larger than zero within a large frequency range,
where the largest value of $F_{a,out}(\omega )$ is about $13.45$ dB at the
central resonance frequency. On the contrary, $F_{b,out}(\omega )$ is zero
over the whole frequency range. This means that we can obtain a squeezed
output field in the right side of the waveguide, but there is no squeezing
of the output field in the left side of the waveguide. Thus, the
unidirectional emission of a squeezed source can be achieved. Notably, we can
readily switch the emission direction of the squeezed field by reversing the
direction of the external biased magnetic field.

In practice, our system will inevitably suffer from the
non-ideal chiral coupling, i.e., $g_{a }\neq0$ and $g_{b }\neq0$, such that
that the completely one-way squeezing emission will be spoiled. In the
presence of non-ideal chiral coupling, we know from Eq. (\ref{eq10}) that the
hybridized mode $A$ will be steered into a squeezed state via the
dissipation of the magnon mode $m$, i.e., both the modes $a$ and $b$ are
thereby squeezed. As a result, the squeezed fields will be emitted in both
directions of the waveguide. With the increase of $g_{b }$, the associated
output squeezing $F_{a,out}(\omega )$ will be reduced, while the output
squeezing $F_{b,out}(\omega )$ is gradually increased. Nonetheless, for
$g_{b}/g_{a}=0.1 $, the value of $F_{a,out}(\omega )$ only has a negligible
reduction, and the maximal value of $F_{b,out}(\omega )$ is about $0.04$ dB [See Fig. 3].
So, our scheme is robust against the imperfect chiral coupling, indicating that
we can still implement a high-performance unidirectional squeezing emitter.

\begin{figure}[tbh]
\centering\includegraphics[width=9cm]{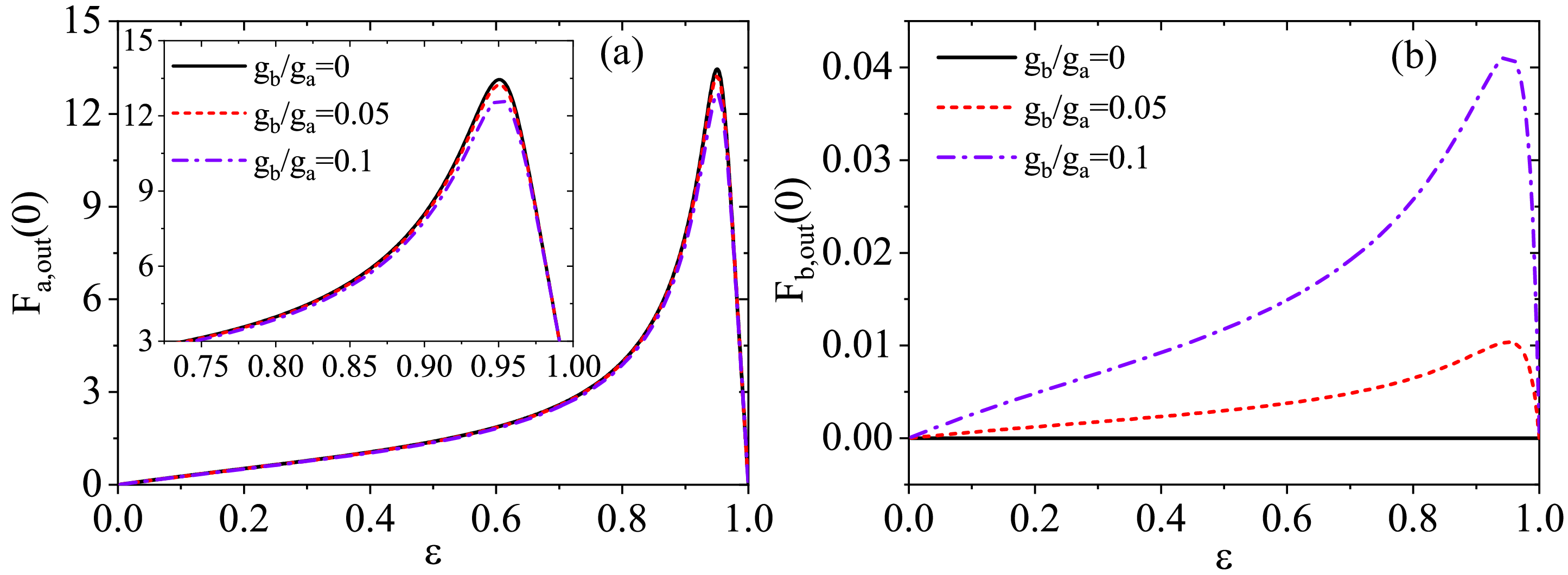}
\caption{(Color online) The squeezing degree $F_{a,out}(0)$ [(a)] and $%
F_{b,out}(0)$ [(b)] versus the parameter $\protect\varepsilon $ under the different
coupling ratios $g_{b}/g_{a}=0$, $0.05$, $0.1$. The other parameters are the
same as in Fig. 3.}
\end{figure}

Finally, we emphasize here that the proposed one-way
emitter with a tunable amount of squeezing can also be generated in our
scheme. Fig. 4 displays the $F_{a,out}(0)$ and $F_{b,out}(0)$ versus the
parameter $\varepsilon $. We can see that $F_{a,out}(0)$ behaves as a
nonmonotonic function of $\varepsilon $, implying that a tunable squeezed
output field can be created by tuning the parameter $\varepsilon $. This can
be implemented by adjusting the Floquet driving parameters $\Omega
_{j}(j=1,2)$.

As presented in Fig. 4(a), there exists an optimum value of $\varepsilon $,
where a maximal amount of output squeezing is produced. To illustrate this
point, we recall that the intra-cavity squeezing depends on the parameter $%
\varepsilon $. Meanwhile, according to the standard input-output relation,
the intra-cavity squeezing determines the output one. So, the output
squeezing is also determined by the parameter $\varepsilon $. Based on Eqs.
(\ref{eq9}) and (\ref{eq10}), we know that the dissipative magnon mode can be used to cool
the cavity mode down to a squeezed state. So, it seems like that the
intra-cavity squeezing will be continuously increased with the increase of $%
\varepsilon $. However, as the parameter $\varepsilon $ increasing, the
effective magnon-photon coupling, which plays the role of cooling, is
gradually decreased in the squeezed representation [See Eq. (\ref{eq10})]. Besides,
the cavity vacuum noise is transformed to the effective thermal noise, which
is also be amplified. As a result, the cavity can not approach a squeezed
vacuum state, which is actually a thermal squeezed state. Therefore, the
competition of these two processes leads to an optimum value of $\varepsilon $
that can generate a peak of output squeezing.

\section{Conclusion}

In summary, we have presented an approach for the unidirectional emission of
a tunable squeezed microwave field in a chiral cavity magnonic system of a
YIG sphere and a torus-shaped cavity. The chirality comes from the selective
coupling between the Kittel mode to one of the two counter-propagating
microwave modes with the same polarization. Under a bichromatic Floquet
driving field to induce sidebands, the unidirectional emission of a tunable
squeezed microwave source can be generated with the help of a dissipative
magnon mode and a waveguide. Moreover, the emission direction can be
conveniently changed by varying the direction of the external applied
magnetic field. With the rapid development of hybrid cavity magnonic system,
our scheme is expected to be implemented in a realistic experiment, which
can stimulate a wide range of quantum technological applications.

\section*{Acknowledgement}

This work was supported by the National Natural Science Foundation of China
(Grant Nos. 11704306 and 12074307).

\bibliography{Ref}

\end{document}